\documentclass[letter,longauth]{aa}  

\usepackage{bm}
\usepackage{graphicx}
\usepackage{txfonts}
\usepackage[colorlinks=true,linkcolor=blue,citecolor=blue,urlcolor=blue]{hyperref}
\usepackage[dvipsnames]{xcolor}
\usepackage[normalem]{ulem}
\usepackage{soul}

\newcommand{\kyr}{\ensuremath{\mathrm{kyr}}}

\newcommand{\mdot}{\ensuremath{\dot{M}}}
\newcommand{\Msun}{\ensuremath{\,M_{\odot}}}	
\newcommand{\pyr}{\ensuremath{\,\mathrm{yr}^{-1}}}	

\newcommand{\vinf}{\ensuremath{\varv_{\infty}}}	
\newcommand{\kms}{\ensuremath{\,~\textrm{km\,s}^{-1}}}	
\newcommand{\teff}{\ensuremath{T_{\mathrm{eff}}}}

\usepackage{xcolor}
\usepackage{ifthen}
\usepackage{subcaption}    
\usepackage{etoolbox}

\newbool{showred}
\setbool{showred}{false} 

\newcommand{\new}[1]{%
  \ifbool{showred}
    {\textcolor{red}{#1}}
    {#1}%
}

\begin{document} 
\authorrunning{
    Larkin et al.}

    \title{EWOCS-V: Is Wd1-72 a recent post-interaction WR+O binary?}

   \author{C. J. K. Larkin\inst{\ref{inst:mpik},\ref{inst:ari},\ref{inst:mpia}}
          \and
             J. Mackey\inst{\ref{inst:dias}}
             \and
             H. Jin\inst{\ref{inst:mpa}}
            \and
          A. A. C. Sander\inst{\ref{inst:ari},\ref{inst:iwr}}
          \and
          B. Reville\inst{\ref{inst:mpik}}
          \and
          K. Anastasopoulou \inst{\ref{inst:inaf_pal}}
           \and
          M. Andersen \inst{\ref{inst:eso}}
          \and
         A. Bayo \inst{\ref{inst:eso}}
         \and
         J. J. Drake \inst{\ref{inst:LM}}
         \and
          E. K. Grebel\inst{\ref{inst:ari}}
         \and
          M. G. Guarcello \inst{\ref{inst:inaf_pal}}
          \and
          T. J. Haworth \inst{\ref{inst:qmul}}
          \and
          V. M. Kalari\inst{\ref{inst:gemini}}
          \and
          R. R. Lefever\inst{\ref{inst:ari}}
          \and
          F. Najarro\inst{\ref{inst:cab}}
          \and
          B. W. Ritchie 
          \inst{\ref{inst:ou}}
           \and
          E. Sabbi
          \inst{\ref{inst:gemini}}
          }

   \institute{{Max-Planck-Institut f\"{u}r Kernphysik, Saupfercheckweg 1, D-69117 Heidelberg, Germany\label{inst:mpik}}\\
            \email{cormac.larkin@mpi-hd.mpg.de}
         \and 
         {Zentrum für Astronomie der Universität Heidelberg, Astronomisches Rechen-Institut, M{\"o}nchhofstr. 12-14, 69120 Heidelberg, Germany\label{inst:ari}}
         \and 
         {Max-Planck-Institut f\"{u}r Astronomie, K\"{o}nigstuhl 17, D-69117 Heidelberg, Germany\label{inst:mpia}}
                 \and
          {Astronomy \& Astrophysics Section, School of Cosmic Physics, Dublin Institute for Advanced Studies, DIAS Dunsink Observatory, Dublin D15 XR2R, Ireland\label{inst:dias}}
            \and 
         {Max Planck Institute for Astrophysics, Karl-Schwarzschild-Str. 1, 85748 Garching, Germany\label{inst:mpa}}
         \and
    {Universit\"at Heidelberg, Interdisziplin\"ares Zentrum f\"ur Wissenschaftliches Rechnen, 69120 Heidelberg, Germany\label{inst:iwr}}
    \and
    {Istituto Nazionale di Astrofisica (INAF) – Osservatorio Astronomico di Palermo, Piazza del Parlamento 1, 90134 Palermo, Italy\label{inst:inaf_pal}}
    \and
    {European Southern Observatory, Karl-Schwarzschild-Strasse 2, 85748 Garching bei M\"unchen, Germany\label{inst:eso}}
    \and
    {Lockheed Martin Solar and Astrophysics Laboratory, 3251 Hanover Street, Palo Alto, CA 94304, USA\label{inst:LM}}
    \and
    {School of Physical and Chemical Sciences, Queen Mary University of London, Mile End, London E1 4NS, UK\label{inst:qmul}}
    \and
    {Gemini Observatory/NSFs NOIRLab, 950 N. Cherry Ave., Tucson, AZ 85719, USA\label{inst:gemini}}
     \and
    {Departamento de Astrofísica, Centro de Astrobiología, CSIC-INTA, Ctra. Torrejón a Ajalvir km 4, E-28850 Torrejón de Ardoz, Spain\label{inst:cab}}
    \and
    {Department of Physics and Astronomy, The Open University, Walton Hall, Milton Keynes MK7 6AA, UK\label{inst:ou}}
                      }

   \date{Received DATE; accepted DATE}

  \abstract{The evolutionary origin of Wolf-Rayet (WR) stars at Solar metallicity is unclear. Single-star evolution from massive O stars, possibly via a Luminous Blue Variable phase, is challenged by binary period distributions of different WR subtypes. Wd1-72 is a WN7b+O binary embedded in the collective wind of the Galactic young massive cluster Westerlund 1 (Wd~1). It is surrounded by highly structured nebulosity, with cometary tails pointing away from Wd~1 and quasi-spherical droplets towards it. In this letter, we demonstrate that this morphology can be qualitatively reproduced by a hydrodynamic simulation of non-conservative Roche Lobe Overflow (RLOF) mass-loss into a cluster wind. Our model is based on a detailed binary evolution track consistent with key known properties of Wd1-72. Our work suggests Wd1-72 could be only $\sim$10~\kyr\ post-RLOF, and the hydrogen-free nature of Wd1-72 favours this being a second or subsequent RLOF episode. Follow-up observations could make Wd1-72 a valuable benchmark for probing mass-loss and mass-transfer in forming gravitational-wave binary-progenitor systems.  
  }


   \maketitle
\nolinenumbers

\section{Introduction}

In the classical single-star scenario proposed by \citet{Conti1975}, massive O stars undergo intense mass-loss (ML), stripping the hydrogen (H) envelope and producing a Wolf-Rayet (WR) star. This may apply to the most massive stars \citep[$M_\text{init} \gtrsim 100\,M_\odot$, e.g.,][]{Sabhahit2022}, but steady winds alone are potentially insufficient to do this for a wider range of initial masses. A brief Luminous Blue Variable (LBV) phase with eruptive ML has therefore been suggested \citep{Langer1994} and is applied in grids of massive-star evolution models \citep[e.g.,][]{Ekstroem2012}. In such evolution models, the resulting WR stars usually begin with a nitrogen-rich surface (WN-type) of a late subtype and then pass gradually to earlier subtypes and eventually to the carbon-rich WC stage \citep[e.g.][]{Groh2014}. However, while there are a few observations of transitional WN/WC and even WN/WO stars \citep[e.g.,][]{ContiMassey1989,Sander2025}, the single-star nature of larger LBV eruptions has been questioned \citep[e.g.,][]{Boffin2016,Hirai2021}.

The majority of massive stars are in multiple systems and will experience some binary interaction in their lifetime \citep[e.g.][]{Sana2012,deMink2014}. Such effects have long been suggested as an alternative path to strip away enough of the outer envelope to form a WR star \citep[e.g.][]{Paczynski1967}. Binary stripping could also break the presumption of self-stripping where WR stars form as (very) late WN subtypes and then consecutively evolve to earlier ones. \citet{Dsilva2023} showed that the Galactic WN binary population have mostly short periods (1-10 d) whereas the WC population distribution peaks at $\sim$$5000$\,d. This implies their formation pathways are different, leaving many open questions about descendants of these systems. This has implications for not only our understanding of WR formation, but also the properties of the progenitors of gravitational waves from black-hole binary mergers \citep[e.g,][]{Higgins2021,Marchant2024}.

The young massive cluster Westerlund 1 (Wd~1) is an ideal laboratory to study massive stellar evolution due to its large, diverse and likely coeval (age~$\sim5.5$~Myr,~\citealt{Castellanos2026}, although this is debated by e.g. \citealt{Beasor2021}) population of massive stars. This includes 24 WR stars \citep[$>$90\% of which are binaries,][]{2006MNRAS.372.1407C,Anastasopoulou2024} as well as the sgB[e] star Wd1-9 \citep{2020A&A...635A.187C}, likely a WR+OB binary system that recently experienced Case-B mass transfer \citep{Anastasopoulou2025}. JWST observations of Wd~1 by the Extended Westerlund 1 and 2 Open Clusters Survey (EWOCS) collaboration show the cluster is surrounded by complex nebulosity \citep{Guarcello2024}---a finding unexpected for a cluster this old because feedback should disperse leftover natal gas on much shorter timescales \citep{PortegiesZwart2010}. 

In the west and south of Wd~1, the nebulosity appears to be embedded in a collective radially expanding cluster wind, driven by the powerful WR stars in the core. In the east however, the cometary-like droplets appear to be centered on Wd1-72, a WN7b + late O supergiant binary system (alias WR A or WR\,77sc) with two periods observed. A 7.63\,d period ($P$) was reported by \citet{Bonanos2007} in optical data and also observed in X-rays by \citet{Anastasopoulou2024}, who found an additional stronger $P$ of 81\,d. The X-ray spectrum of Wd1-72 is very similar to that of Wd1-9 \citep{Clark2008,Anastasopoulou2024}. Wd1-72 is also a binary with both components similar to the Wd1-9 model proposed by \citet{Anastasopoulou2025}, who suggest it as an analogue for the near future of Wd1-9.   

In this letter, the fifth in the EWOCS series, we demonstrate how both key aspects of the qualitative morphology of the nebulosity around Wd1-72 as seen at 11$\mu$m can be reproduced with a hydrodynamic model of intense non-conservative Roche Lobe Overflow (hereafter RLOF) ML into a cluster wind. The ML properties are motivated by a bespoke binary stellar evolution track consistent with the known properties of Wd1-72. We briefly discuss implications for massive binary evolution, and how follow-up observations are required to constrain the recent ML history, and test the possible post-interaction nature of, Wd1-72. Our letter is organised as follows: In Sect. \ref{sec:methods} we describe our evolutionary track. In Sect. \ref{sec:results} we present our simulation and compare it qualitatively with JWST observations. We discuss our findings and present our conclusions in Sect. \ref{sec:discussion}. In the appendices we describe our simulation setup (App. \ref{sec:App_hydro}), show how our evolutionary track compares to the measured properties of Wd1-72 (App. \ref{sec:App_evol}), physical constraints on the eruption and droplet properties disfavouring a giant LBV eruption (App. \ref{sec:App_analytic}), and estimates of dust emission from the dense clumps produced in the simulation (App. \ref{sec:App_torus}).

\section{Methods}
\label{sec:methods}

 We use a newly computed MESA \citep[version 10398,][]{MESA_I,MESA_II,MESA_III,MESA_IV,MESA_V} track with the same physics assumptions as in \citet{Jin2025}, except for the wind mass-loss rate (\mdot, see \citet{Jin2024} for details). In test hydrodynamic simulations the RLOF material was found to disperse after $\lesssim\,$$50$~\kyr, setting an upper limit for the timescale of depletion of surface H after a RLOF event. O stars are expected to retain some surface H after rapid RLOF in an intermediate slow RLOF phase for $\gtrsim\,$$100$\kyr\, \citep{Clark2014}, which would appear as e.g. a WN9h spectral type, so this favours an evolutionary path with multiple RLOF episodes. One model with two RLOF events was identified in the grid of \citet{Jin2025} which was close to the known age, luminosity ($L_{\mathrm{bol}}$) and $P$ of Wd1-72, but was H free after $\sim70$\kyr. This model is at Solar metallicity, has an initial $P$ of 3.98 d, primary (initially more massive) $M_\text{init}=28.18\Msun$ and secondary $M_\text{init}=19.73\Msun$. We reran this evolutionary track with a factor 4.7 increase in the \mdot\ of the primary star starting immediately after the second RLOF event, to approximately match the rate of $9\times10^{-5}\Msun\pyr$ measured from spectra of Wd1-72 by \citet{Rosslowe2015}. This model reaches a surface H abundance of $\sim1\%$ only $\sim$$20$\kyr\ post-RLOF (see left panel of Fig. \ref{fig:mdot_hydro}) and was used in our hydrodynamic simulation described in App. \ref{sec:App_hydro}. We show the evolution of \mdot, stellar wind velocity (\vinf), $P$, $L_{\mathrm{bol}}$ and H surface abundance in the left panel of Fig. \ref{fig:mdot_hydro} and compare this MESA track with Wd1-72 in App. \ref{sec:App_evol}.

\section{Results}
\label{sec:results}

\begin{figure*}
    \centering
    \includegraphics[width=1\linewidth]{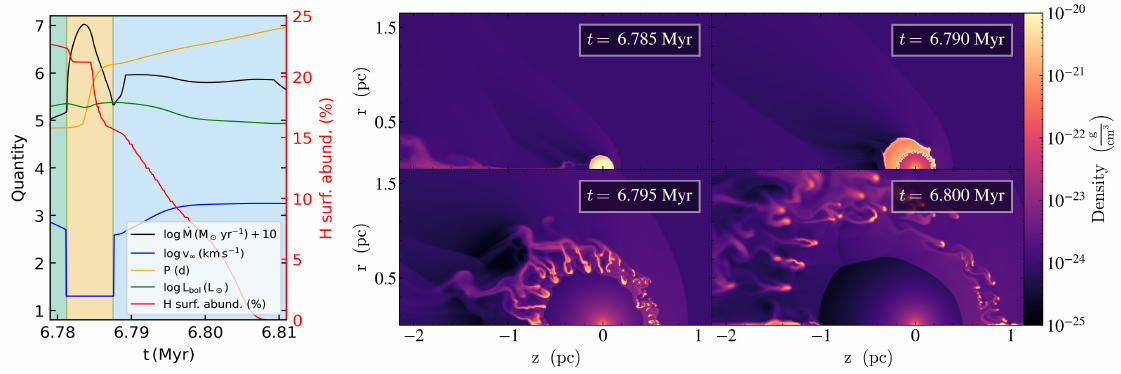}
    \caption{Left: Evolution of \mdot, \vinf, $P$, $L_{\mathrm{bol}}$ and surface H abundance in our MESA track. The green, yellow and blue panels correspond to pre-RLOF, RLOF and WR wind phases. Right: Density slices for four key stages in the evolution of our simulation, with the cluster wind moving in the $-z$ direction.}
    \label{fig:mdot_hydro}
\end{figure*}

We show four key stages in the simulation evolution in the right panel of Fig. \ref{fig:mdot_hydro}. The first panel shows RLOF material filling the pre-RLOF stellar bubble. The second panel shows the WR wind expanding into the RLOF material, with Rayleigh-Taylor (RT) fingers developing. The RLOF material asymmetry arises from the cluster wind. The third panel shows the RLOF material fragmented into ``RT clumps'', with two distinct morphologies. The RT clumps manifest as small, quasi-spheroidal droplets in the $+z$ (upstream) direction in the wake of the ejecta standoff bow shock (see App. \ref{sec:App_analytic}). In the $-z$ direction, the structure is more filamentary. These differences arise due to the cluster wind, as the upstream RT clumps encounter a stagnation region where the flows collide, whereas the downstream RT clumps expand from one side only. In the final panel, the downstream RT clumps have separated into cometary-like droplets with tails. The upstream RT clumps disperse radially from the star on ballistic trajectories, whereas the downstream RT clumps are starting to be carried away in the cluster wind. This fragmentation is dependent on the simulation spatial resolution, with higher resolution leading to more efficient and rapid fragmentation. At this time, for the $+z$ half of the domain we find a total mass in clumps $M_{\mathrm{cl}}=7.3\times10^{-3}$\Msun\ and average clump density $\rho_{\mathrm{cl}} = 8.8\times10^{-22}\mathrm{~g~cm}^{-3}$. For the $-z$ half we obtain $M_{\mathrm{cl}}=3.8\times10^{-2}$\Msun\ and $\rho_{\mathrm{cl}} = 2.9\times10^{-22}\mathrm{~g~cm}^{-3}$. Both halves have average Atwood numbers $\sim0.99$. We exclude any clumps in the region within 0.1 pc in $r$ due to the $z$ axis boundary.

\begin{figure*}
    \centering
    \includegraphics[width=0.9\linewidth]{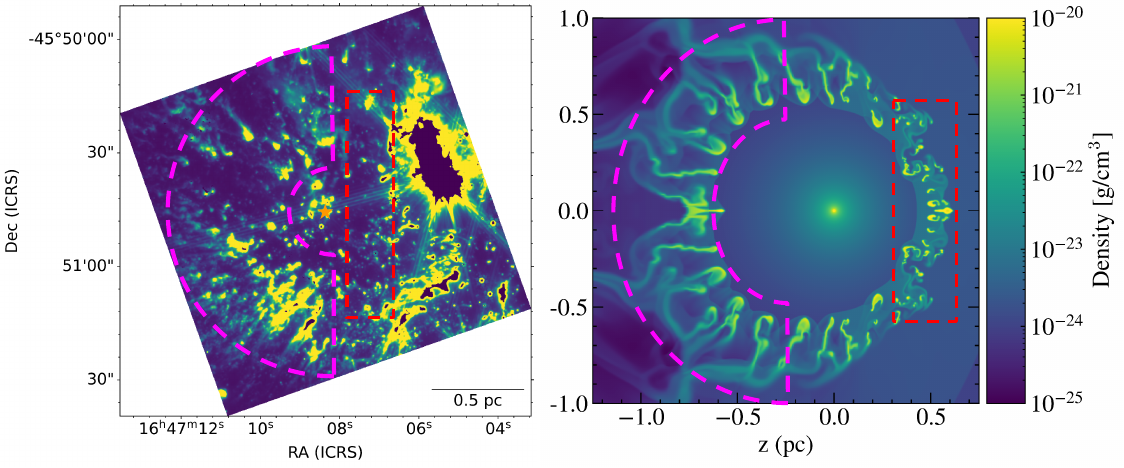}
    \caption{Left: JWST/MIRI F1130W observations towards Wd1-72, which is marked with an orange star. Right: Density slice from our simulation mirrored around the $z$-axis at $t = 6.795$ Myr. We highlight the regions of spheroidal droplets in red, and cometary-like droplets pointing towards Wd1-72 in pink, demonstrating the qualitative correspondence between the data and simulation.}
    \label{fig:morphologycomparison}
\end{figure*}

In Fig. \ref{fig:morphologycomparison} we compare the density field of our simulation with the JWST F1130W data towards Wd1-72. In the direction towards the cluster core we observe quasi-spheroidal droplets in the data, which we also find in our simulation. In the opposite direction, the material around Wd1-72 is structured in droplets with cometary-like tails. These tails point away from the star as opposed to the cluster core (as seen elsewhere in Wd~1), which is also reproduced by our simulation. The main difference is the non-uniformity of the cometary tail distribution in the data, with the upper part of the figure being more sparsely populated than the lower part. We do not observe this kind of spatial variability in our simulation. The material appears to be more filamentary towards the bottom, whereas more distinct droplets with tails are seen towards the top. We observe both of these morphologies in our simulation, but not at the same time. Finally, the material in the observations appears to be a factor $\sim2$ closer to Wd1-72 than in our simulations. This can be at least partially accounted for by the resolution dependence of the fragmentation, and the 2D nature of our simulation, as fragmentation would be more efficient in 3D. These results also depend on our assumptions of the density and velocity of both the stellar and cluster wind, and our assumption of a spherically symmetric outflow.

\section{Discussion \& conclusions}
\label{sec:discussion}
In this letter we demonstrate that the complex nebulosity surrounding Wd1-72 can be explained via intense ML characteristic of non-conservative RLOF being fragmented by a WR wind and carried away in a cluster wind. This scenario should not strongly depend on the specific stellar evolution track, and we stress that the MESA track used here is only one possibility. The WN7 subtype favours this putative RLOF event being a second or subsequent one, as complete H removal in a single RLOF event is unlikely to occur before the droplets are dispersed. Our results are also similar to stellar wind bubble simulations of a WR wind fragmenting and sweeping up a shell of previously ejected red supergiant material \citep[e.g.,][]{GarciaSegura1996a}.  

The most significant simplifying assumption we have made is a spherically symmetric outflow from Wd1-72. While an equatorial outflow would arguably be more realistic given the directional dependence of mass lost via $L_1$ \citep[e.g.,][]{Scherbak2025}, our results would then be sensitive to the (presently unknown) system orientation with respect to the cluster wind. This difference in geometry could account for the non-uniform distribution of material around Wd1-72. We further assumed a fixed RLOF outflow velocity of 20\kms, consistent with values of a few tens of\kms\ expected for close massive binary systems \citep{Scherbak2025}. Given the high $L/M$ and extended radii for the new WR immediately post-mass transfer, which both exceed values predicted for He main sequence stars, we expect an enhanced wind mass-loss rate \citep[e.g.,][]{Sander+2020,Sander2023}, in line with our adjusted MESA model. Finally, we did not consider radiation pressure accelerating the dusty RLOF material, which is expected to be negligible at scales $\gtrsim 0.1$ pc. 

It has been suggested that rapidly rotating WR stars can be produced by LBV eruptions \citep{Vink2011}. However, ejecta nebulae around WR stars are rarely associated with binary central stars \citep{Stock2010} and we demonstrate in App.\,\ref{sec:App_analytic} that a typical giant eruption \mdot\, of order 0.1\,\Msun\pyr\ is incompatible with the ejecta morphology, which requires eruptive \mdot\, to be a few $10^{-4}$\Msun\pyr\ at most. Given the WN7 subtype, this degree of H stripping from a minor eruption ($\lesssim$~0.5~\Msun) is disfavoured.

Additional observations will greatly aid efforts to understand the outflows in Wd~1, especially around Wd1-72 as existing NIRCam JWST data here are badly saturated and affected by diffraction spikes. Integral field spectroscopy would probe the physical conditions, line-of-sight velocities, compositions and inclination angle of the droplets around Wd1-72, testing our proposed evolutionary scenario. An additional epoch of imaging with e.g. JWST/MIRI would provide proper motion measurements of the droplets constraining both the currently unknown transition timescale from sgB[e] to newly-emerged WR binary and Wd~1's collective wind velocity. Finally, multi-epoch medium resolution optical/NIR spectra would probe which of the two X-ray periodicities is that of the binary system. This could also provide the rotational velocity of the secondary, which should be spun up if mass transfer (MT) happened recently.

If Wd1-72 is exiting a phase of recent MT, it suggests that unstable MT occurs in short period WN binaries. Considering that $>$90\% of Wd~1's WR stars are in binaries, it would also add to the growing body of evidence that binary interactions are important for WR evolution at all metallicities. If our proposed scenario is true, then we have found the first short-period WN binary stripped recently enough to have RLOF material that can be quantitatively studied. Given the current lack of observed unstable MT systems \citep{Marchant2024}, further study of Wd1-72 (and Wd1-9, if it is also post-MT) and its surroundings would then provide empirical benchmarks for multi-dimensional models of non-conservative MT outflows, and tests for MT efficiency assumptions used in current and future binary population synthesis codes. Both of these are poorly constrained at present but are very relevant to understand the evolutionary sequence from massive binary to gravitational wave merger event (independent of Wd1-72 and/or Wd1-9 being merger progenitors themselves). A secondary effect here are the stellar wind properties of recent post-MT systems. The difference between the measured and MESA \mdot\ of Wd1-72 suggests that stellar winds could remove more angular momentum over the thermal timescale ($\sim$10~\kyr, as the donor returns to equilibrium), than accounted for in current wind prescriptions. This presents an additional uncertainty in massive binary evolution, ultimately affecting which systems end up as gravitational wave mergers. We conclude that:
\begin{itemize}
    \item The complex morphology of nebulosity in the vicinity of Wd1-72 can be reproduced by non-conservative RLOF mass-loss followed by a Wolf-Rayet stellar wind expanding into a cluster wind.
        \item Follow-up observations of Wd1-72 and the nearby material are needed to test whether this material is from RLOF mass-loss, if the system is a short-period binary, and if the companion has been spun up.
    \item If Wd1-72 and Wd1-9 represent different evolutionary stages of recent post-interaction WN binaries, then further study of them and their RLOF material could help address how mass transfer shapes these systems, improving our understanding of gravitational wave binary progenitor systems.
    
\end{itemize}

\begin{acknowledgements}
 We thank the referee for their constructive feedback which has improved the letter. We thank P.A.\ Crowther, A.\ Gilkis, M.\ Gronke, M.\ Mapelli, F.R.N.\ Schneider and J.S.\ Vink for discussions. CJKL gratefully acknowledges support from the International Max Planck Research School for Astronomy and Cosmic Physics at the University of Heidelberg in the form of an IMPRS PhD fellowship. AACS, RRL and CJKL are supported by the Deutsche Forschungsgemeinschaft (DFG, German Research Foundation) in the form of an Emmy Noether Research Group – Project-ID 445674056 (SA4064/1-1, PI Sander). This project was co-funded by the European Union (Project 101183150 - OCEANS). TJH  acknowledges a Dorothy Hodgkin Fellowship, UKRI guaranteed funding for a Horizon Europe ERC consolidator grant (EP/Y024710/1) and UKRI/STFC grant ST/X000931/1. A.B acknowledges support from the Deutsche Forschungsgemeinschaft (DFG, German Research Foundation) under Germany's Excellence Strategy – EXC 2094 – 390783311. FN gratefully acknowledges support by grant PID2022-137779OB-C41 funded by the Spanish Ministry of Science, Innovation and Universities/State Agency of Research MICIU/AEI/10.13039/501100011033 and by ``ERDF A way of making Europe''. E.S. is supported by the international Gemini Observatory, a program of NSF NOIRLab, which is managed by the Association of Universities for Research in Astronomy (AURA) under a cooperative agreement with the U.S. National Science Foundation, on behalf of the Gemini partnership of Argentina, Brazil, Canada, Chile, the Republic of Korea, and the United States of America. The simulations presented here were performed on the HPC system Raven at the Max Planck Computing and Data Facility. This research has made use of the Astrophysics Data System, funded by NASA under Cooperative Agreement 80NSSC21M00561. This study used these software packages: PION \citep{2021PION}, pypion \citep{GreMac21}, Numpy \citep{HarMilVan20}, matplotlib \citep{Hun07}, yt \citep{TurSmiOis11}, TORUS \citep{Harries2019}.   
\end{acknowledgements}

\bibliographystyle{aa} 
\bibliography{biblio.bib} 

@article{MESA_I,
       author = {{Paxton}, Bill and {Bildsten}, Lars and {Dotter}, Aaron and {Herwig}, Falk and {Lesaffre}, Pierre and {Timmes}, Frank},
        title = "{Modules for Experiments in Stellar Astrophysics (MESA)}",
      journal = {\apjs},
         year = 2011,
        month = jan,
       volume = {192},
       number = {1},
          eid = {3},
        pages = {3},
          doi = {10.1088/0067-0049/192/1/3},
archivePrefix = {arXiv},
       eprint = {1009.1622},
 primaryClass = {astro-ph.SR},
       adsurl = {https://ui.adsabs.harvard.edu/abs/2011ApJS..192....3P}
}

@article{MESA_II,
       author = {{Paxton}, Bill and {Cantiello}, Matteo and {Arras}, Phil and {Bildsten}, Lars and {Brown}, Edward F. and {Dotter}, Aaron and {Mankovich}, Christopher and {Montgomery}, M.~H. and {Stello}, Dennis and {Timmes}, F.~X. and {Townsend}, Richard},
        title = "{Modules for Experiments in Stellar Astrophysics (MESA): Planets, Oscillations, Rotation, and Massive Stars}",
      journal = {\apjs},
         year = 2013,
        month = sep,
       volume = {208},
       number = {1},
          eid = {4},
        pages = {4},
          doi = {10.1088/0067-0049/208/1/4},
archivePrefix = {arXiv},
       eprint = {1301.0319},
 primaryClass = {astro-ph.SR},
       adsurl = {https://ui.adsabs.harvard.edu/abs/2013ApJS..208....4P}
}

@article{MESA_III,
       author = {{Paxton}, Bill and {Marchant}, Pablo and {Schwab}, Josiah and {Bauer}, Evan B. and {Bildsten}, Lars and {Cantiello}, Matteo and {Dessart}, Luc and {Farmer}, R. and {Hu}, H. and {Langer}, N. and {Townsend}, R.~H.~D. and {Townsley}, Dean M. and {Timmes}, F.~X.},
        title = "{Modules for Experiments in Stellar Astrophysics (MESA): Binaries, Pulsations, and Explosions}",
      journal = {\apjs},
         year = 2015,
        month = sep,
       volume = {220},
       number = {1},
          eid = {15},
        pages = {15},
          doi = {10.1088/0067-0049/220/1/15},
archivePrefix = {arXiv},
       eprint = {1506.03146},
 primaryClass = {astro-ph.SR},
       adsurl = {https://ui.adsabs.harvard.edu/abs/2015ApJS..220...15P}
}

@ARTICLE{MESA_IV,
       author = {{Paxton}, Bill and {Schwab}, Josiah and {Bauer}, Evan B. and {Bildsten}, Lars and {Blinnikov}, Sergei and {Duffell}, Paul and {Farmer}, R. and {Goldberg}, Jared A. and {Marchant}, Pablo and {Sorokina}, Elena and {Thoul}, Anne and {Townsend}, Richard H.~D. and {Timmes}, F.~X.},
        title = "{Modules for Experiments in Stellar Astrophysics (MESA): Convective Boundaries, Element Diffusion, and Massive Star Explosions}",
      journal = {\apjs},
         year = 2018,
        month = feb,
       volume = {234},
       number = {2},
          eid = {34},
        pages = {34},
          doi = {10.3847/1538-4365/aaa5a8},
archivePrefix = {arXiv},
       eprint = {1710.08424},
 primaryClass = {astro-ph.SR},
       adsurl = {https://ui.adsabs.harvard.edu/abs/2018ApJS..234...34P}
}

@article{MESA_V,
       author = {{Paxton}, Bill and {Smolec}, R. and {Schwab}, Josiah and {Gautschy}, A. and {Bildsten}, Lars and {Cantiello}, Matteo and {Dotter}, Aaron and {Farmer}, R. and {Goldberg}, Jared A. and {Jermyn}, Adam S. and {Kanbur}, S.~M. and {Marchant}, Pablo and {Thoul}, Anne and {Townsend}, Richard H.~D. and {Wolf}, William M. and {Zhang}, Michael and {Timmes}, F.~X.},
        title = "{Modules for Experiments in Stellar Astrophysics (MESA): Pulsating Variable Stars, Rotation, Convective Boundaries, and Energy Conservation}",
      journal = {\apjs},
         year = 2019,
        month = jul,
       volume = {243},
       number = {1},
          eid = {10},
        pages = {10},
          doi = {10.3847/1538-4365/ab2241},
archivePrefix = {arXiv},
       eprint = {1903.01426},
 primaryClass = {astro-ph.SR},
       adsurl = {https://ui.adsabs.harvard.edu/abs/2019ApJS..243...10P}
}

@ARTICLE{2021PION,
	author = {{Mackey}, Jonathan and {Green}, Samuel and {Moutzouri}, Maria and {Haworth}, Thomas J. and {Kavanagh}, Robert D. and {Zargaryan}, Davit and {Celeste}, Maggie},
	title = "{PION: simulating bow shocks and circumstellar nebulae}",
	journal = {\mnras},
	year = 2021,
	month = jun,
	volume = {504},
	number = {1},
	pages = {983-1008},
	doi = {10.1093/mnras/stab781},
	archivePrefix = {arXiv},
	eprint = {2103.07555},
	primaryClass = {astro-ph.GA},
	adsurl = {https://ui.adsabs.harvard.edu/abs/2021MNRAS.504..983M}
}

@ARTICLE{Sander+2020,
       author = {{Sander}, Andreas A.~C. and {Vink}, J.~S. and {Hamann}, W. -R.},
        title = "{Driving classical Wolf-Rayet winds: a {\ensuremath{\Gamma}}- and Z-dependent mass-loss}",
      journal = {\mnras},
         year = 2020,
        month = jan,
       volume = {491},
       number = {3},
        pages = {4406-4425},
          doi = {10.1093/mnras/stz3064},
archivePrefix = {arXiv},
       eprint = {1910.12886},
 primaryClass = {astro-ph.SR},
       adsurl = {https://ui.adsabs.harvard.edu/abs/2020MNRAS.491.4406S}
}

@ARTICLE{Green2019,
       author = {{Green}, Samuel and {Mackey}, Jonathan and {Haworth}, Thomas J. and {Gvaramadze}, Vasilii V. and {Duffy}, Peter},
        title = "{Thermal emission from bow shocks. I. 2D hydrodynamic models of the Bubble Nebula}",
      journal = {\aap},
         year = 2019,
        month = may,
       volume = {625},
          eid = {A4},
        pages = {A4},
          doi = {10.1051/0004-6361/201834832},
archivePrefix = {arXiv},
       eprint = {1903.05505},
 primaryClass = {astro-ph.GA},
       adsurl = {https://ui.adsabs.harvard.edu/abs/2019A&A...625A...4G}
}

@ARTICLE{Sana2012,
       author = {{Sana}, H. and {de Mink}, S.~E. and {de Koter}, A. and {Langer}, N. and {Evans}, C.~J. and {Gieles}, M. and {Gosset}, E. and {Izzard}, R.~G. and {Le Bouquin}, J. -B. and {Schneider}, F.~R.~N.},
        title = "{Binary Interaction Dominates the Evolution of Massive Stars}",
      journal = {Science},
         year = 2012,
        month = jul,
       volume = {337},
       number = {6093},
        pages = {444},
          doi = {10.1126/science.1223344},
archivePrefix = {arXiv},
       eprint = {1207.6397},
 primaryClass = {astro-ph.SR},
       adsurl = {https://ui.adsabs.harvard.edu/abs/2012Sci...337..444S}
}

@ARTICLE{GarciaSegura1996a,
       author = {{Garcia-Segura}, G. and {Langer}, N. and {Mac Low}, M. -M.},
        title = "{The hydrodynamic evolution of circumstellar gas around massive stars. II. The impact of the time sequence O star -> RSG -> WR star.}",
      journal = {\aap},
         year = 1996,
        month = dec,
       volume = {316},
        pages = {133-146},
       adsurl = {https://ui.adsabs.harvard.edu/abs/1996A&A...316..133G}
}

@software{GreMac21,
       author = {{Green}, Samuel and {Mackey}, Jonathan},
        title = "{PyPion: Post-processing code for PION simulation data}",
 howpublished = {Astrophysics Source Code Library, record ascl:2103.026},
         year = 2021,
        month = mar,
          eid = {ascl:2103.026},
       adsurl = {https://ui.adsabs.harvard.edu/abs/2021ascl.soft03026G}
}

@Article{HarMilVan20,
 title         = {Array programming with {NumPy}},
 author        = {Charles R. Harris and K. Jarrod Millman and St{'{e}}fan J.
                 van der Walt and Ralf Gommers and Pauli Virtanen and David
                 Cournapeau and Eric Wieser and Julian Taylor and Sebastian
                 Berg and Nathaniel J. Smith and Robert Kern and Matti Picus
                 and Stephan Hoyer and Marten H. van Kerkwijk and Matthew
                 Brett and Allan Haldane and Jaime Fern{'{a}}ndez del
                 R{'{\i}}o and Mark Wiebe and Pearu Peterson and Pierre
                 G{'{e}}rard-Marchant and Kevin Sheppard and Tyler Reddy and
                 Warren Weckesser and Hameer Abbasi and Christoph Gohlke and
                 Travis E. Oliphant},
 year          = {2020},
 month         = sep,
 journal       = {Nature},
 volume        = {585},
 number        = {7825},
 pages         = {357--362},
 doi           = {10.1038/s41586-020-2649-2},
 publisher     = {Springer Science and Business Media {LLC}},
 url           = {https://doi.org/10.1038/s41586-020-2649-2}
}

@Article{Hun07,
  Author    = {Hunter, J. D.},
  Title     = {Matplotlib: A 2D graphics environment},
  Journal   = {Computing in Science \& Engineering},
  Volume    = {9},
  Number    = {3},
  Pages     = {90--95},
  publisher = {IEEE COMPUTER SOC},
  doi       = {10.1109/MCSE.2007.55},
  year      = 2007
}

@ARTICLE{TurSmiOis11,
   author = {{Turk}, M.~J. and {Smith}, B.~D. and {Oishi}, J.~S. and {Skory}, S. and
     {Skillman}, S.~W. and {Abel}, T. and {Norman}, M.~L.},
    title = "{yt: A Multi-code Analysis Toolkit for Astrophysical Simulation Data}",
  journal = {The Astrophysical Journal Supplement Series},
archivePrefix = "arXiv",
   eprint = {1011.3514},
 primaryClass = "astro-ph.IM",
     year = 2011,
    month = jan,
   volume = 192,
      eid = {9},
    pages = {9},
      doi = {10.1088/0067-0049/192/1/9},
   adsurl = {https://ui.adsabs.harvard.edu/abs/2011ApJS..192....9T}
}

@ARTICLE{Guarcello2024,
       author = {{Guarcello}, M.~G. and {Almendros-Abad}, V. and {Lovell}, J.~B. and {Monsch}, K. and {Mu{\v{z}}i{\'c}}, K. and {Mart{\'\i}nez-Galarza}, J.~R. and {Drake}, J.~J. and {Anastasopoulou}, K. and {Andersen}, M. and {Argiroffi}, C. and {Bayo}, A. and {Bonito}, R. and {Capela}, D. and {Damiani}, F. and {Gennaro}, M. and {Ginsburg}, A. and {Grebel}, E.~K. and {Hora}, J.~L. and {Moraux}, E. and {Najarro}, F. and {Negueruela}, I. and {Prisinzano}, L. and {Richardson}, N.~D. and {Ritchie}, B. and {Robberto}, M. and {Rom}, T. and {Sabbi}, E. and {Sciortino}, S. and {Umana}, G. and {Winter}, A. and {Wright}, N.~J. and {Zeidler}, P.},
        title = "{EWOCS-III: JWST observations of the supermassive star cluster Westerlund 1}",
      journal = {\aap},
         year = 2025,
        month = jan,
       volume = {693},
          eid = {A120},
        pages = {A120},
          doi = {10.1051/0004-6361/202452150},
archivePrefix = {arXiv},
       eprint = {2411.13051},
 primaryClass = {astro-ph.SR},
       adsurl = {https://ui.adsabs.harvard.edu/abs/2025A&A...693A.120G}
}

@ARTICLE{Harries2019,
       author = {{Harries}, T.~J. and {Haworth}, T.~J. and {Acreman}, D. and {Ali}, A. and {Douglas}, T.},
        title = "{The TORUS radiation transfer code}",
      journal = {Astronomy and Computing},
         year = 2019,
        month = apr,
       volume = {27},
          eid = {63},
        pages = {63},
          doi = {10.1016/j.ascom.2019.03.002},
archivePrefix = {arXiv},
       eprint = {1903.06672},
 primaryClass = {astro-ph.SR},
       adsurl = {https://ui.adsabs.harvard.edu/abs/2019A&C....27...63H}
}

@ARTICLE{Dra03,
   author = {{Draine}, B.~T.},
    title = "{Interstellar Dust Grains}",
  journal = {\araa},
   eprint = {astro-ph/0304489},
     year = 2003,
   volume = 41,
    pages = {241-289},
      doi = {10.1146/annurev.astro.41.011802.094840},
   adsurl = {http://adsabs.harvard.edu/abs/2003ARA%26A..41..241D}
}

@ARTICLE{2006MNRAS.372.1407C,
       author = {{Crowther}, Paul A. and {Hadfield}, L.~J. and {Clark}, J.~S. and {Negueruela}, I. and {Vacca}, W.~D.},
        title = "{A census of the Wolf-Rayet content in Westerlund 1 from near-infrared imaging and spectroscopy}",
      journal = {\mnras},
         year = 2006,
        month = nov,
       volume = {372},
       number = {3},
        pages = {1407-1424},
          doi = {10.1111/j.1365-2966.2006.10952.x},
archivePrefix = {arXiv},
       eprint = {astro-ph/0608356},
 primaryClass = {astro-ph},
       adsurl = {https://ui.adsabs.harvard.edu/abs/2006MNRAS.372.1407C}
}

@ARTICLE{2020A&A...635A.187C,
       author = {{Clark}, J.~S. and {Ritchie}, B.~W. and {Negueruela}, I.},
        title = "{A VLT/FLAMES survey for massive binaries in Westerlund 1. VII. Cluster census}",
      journal = {\aap},
         year = 2020,
        month = mar,
       volume = {635},
          eid = {A187},
        pages = {A187},
          doi = {10.1051/0004-6361/201935903},
archivePrefix = {arXiv},
       eprint = {1908.05616},
 primaryClass = {astro-ph.SR},
       adsurl = {https://ui.adsabs.harvard.edu/abs/2020A&A...635A.187C}
}

@ARTICLE{2018MNRAS.477L..55A,
       author = {{Andrews}, H. and {Fenech}, D. and {Prinja}, R.~K. and {Clark}, J.~S. and {Hindson}, L.},
        title = "{Asymmetric ejecta of cool supergiants and hypergiants in the massive cluster Westerlund 1}",
      journal = {\mnras},
         year = 2018,
        month = jun,
       volume = {477},
       number = {1},
        pages = {L55-L59},
          doi = {10.1093/mnrasl/sly046},
archivePrefix = {arXiv},
       eprint = {1803.07008},
 primaryClass = {astro-ph.SR},
       adsurl = {https://ui.adsabs.harvard.edu/abs/2018MNRAS.477L..55A}
}

@ARTICLE{Clark2008,
       author = {{Clark}, J.~S. and {Muno}, M.~P. and {Negueruela}, I. and {Dougherty}, S.~M. and {Crowther}, P.~A. and {Goodwin}, S.~P. and {de Grijs}, R.},
        title = "{Unveiling the X-ray point source population of the Young Massive Cluster Westerlund 1}",
      journal = {\aap},
         year = 2008,
        month = jan,
       volume = {477},
       number = {1},
        pages = {147-163},
          doi = {10.1051/0004-6361:20077186},
       adsurl = {https://ui.adsabs.harvard.edu/abs/2008A&A...477..147C}
}

@ARTICLE{PortegiesZwart2010,
       author = {{Portegies Zwart}, Simon F. and {McMillan}, Stephen L.~W. and {Gieles}, Mark},
        title = "{Young Massive Star Clusters}",
      journal = {\araa},
         year = 2010,
        month = sep,
       volume = {48},
        pages = {431-493},
          doi = {10.1146/annurev-astro-081309-130834},
archivePrefix = {arXiv},
       eprint = {1002.1961},
 primaryClass = {astro-ph.GA},
       adsurl = {https://ui.adsabs.harvard.edu/abs/2010ARA&A..48..431P}
}

@ARTICLE{Larkin2025a,
       author = {{Larkin}, C.~J.~K. and {Mackey}, J. and {Haworth}, T.~J. and {Sander}, A.~A.~C.},
        title = "{Investigating dusty red supergiant outflows in Westerlund 1 with 3D hydrodynamic simulations}",
      journal = {\aap},
         year = 2025,
        month = aug,
       volume = {700},
          eid = {A60},
        pages = {A60},
          doi = {10.1051/0004-6361/202554334},
archivePrefix = {arXiv},
       eprint = {2503.01272},
 primaryClass = {astro-ph.SR},
       adsurl = {https://ui.adsabs.harvard.edu/abs/2025A&A...700A..60L}
}

@ARTICLE{Sander2023,
       author = {{Sander}, A.~A.~C. and {Lefever}, R.~R. and {Poniatowski}, L.~G. and {Ramachandran}, V. and {Sabhahit}, G.~N. and {Vink}, J.~S.},
        title = "{The temperature dependency of Wolf-Rayet-type mass loss. An exploratory study for winds launched by the hot iron bump}",
      journal = {\aap},
         year = 2023,
        month = feb,
       volume = {670},
          eid = {A83},
        pages = {A83},
          doi = {10.1051/0004-6361/202245110},
archivePrefix = {arXiv},
       eprint = {2301.01785},
 primaryClass = {astro-ph.SR},
       adsurl = {https://ui.adsabs.harvard.edu/abs/2023A&A...670A..83S}
}

@ARTICLE{Sabhahit2022,
       author = {{Sabhahit}, Gautham N. and {Vink}, Jorick S. and {Higgins}, Erin R. and {Sander}, Andreas A.~C.},
        title = "{Mass-loss implementation and temperature evolution of very massive stars}",
      journal = {\mnras},
         year = 2022,
        month = aug,
       volume = {514},
       number = {3},
        pages = {3736-3753},
          doi = {10.1093/mnras/stac1410},
archivePrefix = {arXiv},
       eprint = {2205.09125},
 primaryClass = {astro-ph.SR},
       adsurl = {https://ui.adsabs.harvard.edu/abs/2022MNRAS.514.3736S}
}

@ARTICLE{Anastasopoulou2025,
       author = {{Anastasopoulou}, K. and {Guarcello}, M.~G. and {Drake}, J.~J. and {Ritchie}, B. and {De Becker}, M. and {Bayo}, A. and {Najarro}, F. and {Negueruela}, I. and {Sciortino}, S. and {Flaccomio}, E. and {Castellanos}, R. and {Albacete-Colombo}, J.~F. and {Andersen}, M. and {Damiani}, F. and {Fraschetti}, F. and {Gennaro}, M. and {Gunderson}, S.~J. and {Larkin}, C.~J.~K. and {Mackey}, J. and {Moffat}, A.~F.~J. and {Pradhan}, P. and {Saracino}, S. and {Stevens}, I.~R. and {Weigelt}, G.},
        title = "{EWOCS-IV: 1Ms ACIS Chandra observation of the supergiant B[e] star Wd1-9}",
      journal = {\aap},
         year = 2025,
        month = sep,
       volume = {701},
          eid = {A138},
        pages = {A138},
          doi = {10.1051/0004-6361/202555305},
archivePrefix = {arXiv},
       eprint = {2507.17816},
 primaryClass = {astro-ph.HE},
       adsurl = {https://ui.adsabs.harvard.edu/abs/2025A&A...701A.138A}
}

@ARTICLE{Anastasopoulou2024,
       author = {{Anastasopoulou}, K. and {Guarcello}, M.~G. and {Flaccomio}, E. and {Sciortino}, S. and {Benatti}, S. and {De Becker}, M. and {Wright}, N.~J. and {Drake}, J.~J. and {Albacete-Colombo}, J.~F. and {Andersen}, M. and {Argiroffi}, C. and {Bayo}, A. and {Castellanos}, R. and {Gennaro}, M. and {Grebel}, E.~K. and {Miceli}, M. and {Najarro}, F. and {Negueruela}, I. and {Prisinzano}, L. and {Ritchie}, B. and {Robberto}, M. and {Sabbi}, E. and {Zeidler}, P.},
        title = "{EWOCS-II: X-ray properties of the Wolf{\textendash}Rayet stars in the young Galactic super star cluster Westerlund 1}",
      journal = {\aap},
         year = 2024,
        month = oct,
       volume = {690},
          eid = {A25},
        pages = {A25},
          doi = {10.1051/0004-6361/202348914},
archivePrefix = {arXiv},
       eprint = {2408.11087},
 primaryClass = {astro-ph.HE},
       adsurl = {https://ui.adsabs.harvard.edu/abs/2024A&A...690A..25A}
}

@ARTICLE{Marchant2024,
       author = {{Marchant}, Pablo and {Bodensteiner}, Julia},
        title = "{The Evolution of Massive Binary Stars}",
      journal = {\araa},
         year = 2024,
        month = sep,
       volume = {62},
       number = {1},
        pages = {21-61},
          doi = {10.1146/annurev-astro-052722-105936},
archivePrefix = {arXiv},
       eprint = {2311.01865},
 primaryClass = {astro-ph.SR},
       adsurl = {https://ui.adsabs.harvard.edu/abs/2024ARA&A..62...21M}
}

@ARTICLE{Dsilva2023,
       author = {{Dsilva}, K. and {Shenar}, T. and {Sana}, H. and {Marchant}, P.},
        title = "{A spectroscopic multiplicity survey of Galactic Wolf-Rayet stars . III. The northern late-type nitrogen-rich sample}",
      journal = {\aap},
         year = 2023,
        month = jun,
       volume = {674},
          eid = {A88},
        pages = {A88},
          doi = {10.1051/0004-6361/202244308},
archivePrefix = {arXiv},
       eprint = {2212.06927},
 primaryClass = {astro-ph.SR},
       adsurl = {https://ui.adsabs.harvard.edu/abs/2023A&A...674A..88D}
}

@ARTICLE{Higgins2021,
       author = {{Higgins}, E.~R. and {Sander}, A.~A.~C. and {Vink}, J.~S. and {Hirschi}, R.},
        title = "{Evolution of Wolf-Rayet stars as black hole progenitors}",
      journal = {\mnras},
         year = 2021,
        month = aug,
       volume = {505},
       number = {4},
        pages = {4874-4889},
          doi = {10.1093/mnras/stab1548},
archivePrefix = {arXiv},
       eprint = {2105.12139},
 primaryClass = {astro-ph.SR},
       adsurl = {https://ui.adsabs.harvard.edu/abs/2021MNRAS.505.4874H}
}

@ARTICLE{Conti1975,
       author = {{Conti}, P.~S.},
        title = "{On the relationship between Of and WR stars.}",
      journal = {Memoires of the Societe Royale des Sciences de Liege},
         year = 1975,
        month = jan,
       volume = {9},
        pages = {193-212},
       adsurl = {https://ui.adsabs.harvard.edu/abs/1975MSRSL...9..193C}
}

@PHDTHESIS{Rosslowe2015,
       author = {{Rosslowe}, Christopher},
        title = "{Physical Properties of Wolf-Rayet Stars at Infra-red Wavelengths}",
       school = {University of Sheffield, UK},
         year = 2015,
        month = oct,
       adsurl = {https://ui.adsabs.harvard.edu/abs/2016PhDT.......198R}
}

@ARTICLE{ContiMassey1989,
       author = {{Conti}, Peter S. and {Massey}, Philip},
        title = "{Spectroscopic Studies of Wolf-Rayet Stars. IV. Optical Spectrophotometry of the Emission Lines in Galactic and Large Magellanic Cloud Stars}",
      journal = {\apj},
         year = 1989,
        month = feb,
       volume = {337},
        pages = {251},
          doi = {10.1086/167101},
       adsurl = {https://ui.adsabs.harvard.edu/abs/1989ApJ...337..251C}
}

@ARTICLE{Langer1994,
       author = {{Langer}, N. and {Hamann}, W. -R. and {Lennon}, M. and {Najarro}, F. and {Pauldrach}, A.~W.~A. and {Puls}, J.},
        title = "{Towards an understanding of very massive stars. A new evolutionary scenario relating O stars, LBVs and Wolf-Rayet stars.}",
      journal = {\aap},
         year = 1994,
        month = oct,
       volume = {290},
        pages = {819-833},
       adsurl = {https://ui.adsabs.harvard.edu/abs/1994A&A...290..819L}
}

@ARTICLE{Ekstroem2012,
       author = {{Ekstr{\"o}m}, S. and {Georgy}, C. and {Eggenberger}, P. and {Meynet}, G. and {Mowlavi}, N. and {Wyttenbach}, A. and {Granada}, A. and {Decressin}, T. and {Hirschi}, R. and {Frischknecht}, U. and {Charbonnel}, C. and {Maeder}, A.},
        title = "{Grids of stellar models with rotation. I. Models from 0.8 to 120 M$_{{\ensuremath{\odot}}}$ at solar metallicity (Z = 0.014)}",
      journal = {\aap},
         year = 2012,
        month = jan,
       volume = {537},
          eid = {A146},
        pages = {A146},
          doi = {10.1051/0004-6361/201117751},
archivePrefix = {arXiv},
       eprint = {1110.5049},
 primaryClass = {astro-ph.SR},
       adsurl = {https://ui.adsabs.harvard.edu/abs/2012A&A...537A.146E}
}

@ARTICLE{Groh2014,
       author = {{Groh}, Jose H. and {Meynet}, Georges and {Ekstr{\"o}m}, Sylvia and {Georgy}, Cyril},
        title = "{The evolution of massive stars and their spectra. I. A non-rotating 60 M$_{{\ensuremath{\odot}}}$ star from the zero-age main sequence to the pre-supernova stage}",
      journal = {\aap},
         year = 2014,
        month = apr,
       volume = {564},
          eid = {A30},
        pages = {A30},
          doi = {10.1051/0004-6361/201322573},
archivePrefix = {arXiv},
       eprint = {1401.7322},
 primaryClass = {astro-ph.SR},
       adsurl = {https://ui.adsabs.harvard.edu/abs/2014A&A...564A..30G}
}

@ARTICLE{Boffin2016,
       author = {{Boffin}, Henri M.~J. and {Rivinius}, Thomas and {M{\'e}rand}, Antoine and {Mehner}, Andrea and {LeBouquin}, Jean-Baptiste and {Pourbaix}, Dimitri and {de Wit}, Willem-Jan and {Martayan}, Christophe and {Guieu}, Sylvain},
        title = "{The luminous blue variable HR Carinae has a partner. Discovery of a companion with the VLTI}",
      journal = {\aap},
         year = 2016,
        month = sep,
       volume = {593},
          eid = {A90},
        pages = {A90},
          doi = {10.1051/0004-6361/201629127},
archivePrefix = {arXiv},
       eprint = {1607.07724},
 primaryClass = {astro-ph.SR},
       adsurl = {https://ui.adsabs.harvard.edu/abs/2016A&A...593A..90B}
}

@ARTICLE{Hirai2021,
       author = {{Hirai}, Ryosuke and {Podsiadlowski}, Philipp and {Owocki}, Stanley P. and {Schneider}, Fabian R.~N. and {Smith}, Nathan},
        title = "{Simulating the formation of {\ensuremath{\eta}} Carinae's surrounding nebula through unstable triple evolution and stellar merger-induced eruption}",
      journal = {\mnras},
         year = 2021,
        month = may,
       volume = {503},
       number = {3},
        pages = {4276-4296},
          doi = {10.1093/mnras/stab571},
archivePrefix = {arXiv},
       eprint = {2011.12434},
 primaryClass = {astro-ph.SR},
       adsurl = {https://ui.adsabs.harvard.edu/abs/2021MNRAS.503.4276H}
}

@ARTICLE{Sander2025,
       author = {{Sander}, Andreas A.~C. and {Lefever}, Roel R. and {Josiek}, Joris and {Higgins}, Erin R. and {Hirschi}, Raphael and {Oskinova}, Lidia M. and {Pauli}, Daniel and {Pritzkuleit}, Max and {Gallagher}, John S. and {Hamann}, Wolf-Rainer and {Mandel}, Ilya and {Ramachandran}, Varsha and {Shenar}, Tomer and {Todt}, Helge and {Vink}, Jorick S.},
        title = "{Discovery of a transitional type of evolved massive star with a hard ionizing flux}",
      journal = {Nature Astronomy},
         year = 2026,
        month = feb,
       volume = {10},
        pages = {290-305},
          doi = {10.1038/s41550-025-02719-z},
archivePrefix = {arXiv},
       eprint = {2508.18410},
 primaryClass = {astro-ph.SR},
       adsurl = {https://ui.adsabs.harvard.edu/abs/2026NatAs..10..290S}
}

@ARTICLE{Vink2011,
       author = {{Vink}, J.~S. and {Gr{\"a}fener}, G. and {Harries}, T.~J.},
        title = "{In pursuit of gamma-ray burst progenitors: the identification of a sub-population of rotating Wolf-Rayet stars}",
      journal = {\aap},
         year = 2011,
        month = dec,
       volume = {536},
          eid = {L10},
        pages = {L10},
          doi = {10.1051/0004-6361/201118197},
archivePrefix = {arXiv},
       eprint = {1111.5806},
 primaryClass = {astro-ph.SR},
       adsurl = {https://ui.adsabs.harvard.edu/abs/2011A&A...536L..10V}
}

@ARTICLE{Stock2010,
       author = {{Stock}, D.~J. and {Barlow}, M.~J.},
        title = "{A search for ejecta nebulae around Wolf-Rayet stars using the SHS H{\ensuremath{\alpha}} survey}",
      journal = {\mnras},
         year = 2010,
        month = dec,
       volume = {409},
       number = {4},
        pages = {1429-1440},
          doi = {10.1111/j.1365-2966.2010.17124.x},
archivePrefix = {arXiv},
       eprint = {1006.0625},
 primaryClass = {astro-ph.GA},
       adsurl = {https://ui.adsabs.harvard.edu/abs/2010MNRAS.409.1429S}
}

@ARTICLE{Bonanos2007,
       author = {{Bonanos}, Alceste Z.},
        title = "{Variability of Young Massive Stars in the Galactic Super Star Cluster Westerlund 1}",
      journal = {\aj},
         year = 2007,
        month = jun,
       volume = {133},
       number = {6},
        pages = {2696-2708},
          doi = {10.1086/518093},
archivePrefix = {arXiv},
       eprint = {astro-ph/0702614},
 primaryClass = {astro-ph},
       adsurl = {https://ui.adsabs.harvard.edu/abs/2007AJ....133.2696B}
}

@ARTICLE{Larkin2025b,
       author = {{Larkin}, C.~J.~K. and {Hawcroft}, C. and {Mackey}, J. and {Lefever}, R.~R. and {H{\"a}rer}, L. and {Sander}, A.~A.~C.},
        title = "{Mass-loading of outflows from evolving young massive clusters}",
      journal = {\aap},
         year = 2025,
        month = nov,
       volume = {703},
          eid = {L14},
        pages = {L14},
          doi = {10.1051/0004-6361/202557130},
archivePrefix = {arXiv},
       eprint = {2510.06100},
 primaryClass = {astro-ph.GA},
       adsurl = {https://ui.adsabs.harvard.edu/abs/2025A&A...703L..14L}
}

@ARTICLE{Paczynski1967,
       author = {{Paczy{\'n}ski}, B.},
        title = "{Evolution of Close Binaries. V. The Evolution of Massive Binaries and the Formation of the Wolf-Rayet Stars}",
      journal = {\actaa},
         year = 1967,
        month = jan,
       volume = {17},
        pages = {355},
       adsurl = {https://ui.adsabs.harvard.edu/abs/1967AcA....17..355P}
}

@ARTICLE{Castellanos2026,
       author = {{Castellanos}, R. and {Najarro}, F. and {Garcia}, M. and {Negueruela}, I. and {Patrick}, L.~R. and {Ritchie}, B. and {Guarcello}, M.~G. and {Shenar}, T. and {Evans}, C. and {Prinja}, R. and {Fenech}, D.},
        title = "{First spectroscopic identification of the main sequence in Westerlund 1}",
      journal = {\aap},
         year = 2026,
        month = feb,
          eid = {arXiv:2602.24218},
        pages = {arXiv:2602.24218},
        doi   = {10.1051/0004-6361/202558099},
archivePrefix = {arXiv},
       eprint = {2602.24218},
 primaryClass = {astro-ph.GA},
       adsurl = {https://ui.adsabs.harvard.edu/abs/2026arXiv260224218C}
}

@ARTICLE{Beasor2021,
       author = {{Beasor}, Emma R. and {Davies}, Ben and {Smith}, Nathan and {Gehrz}, Robert D. and {Figer}, Donald F.},
        title = "{The Age of Westerlund 1 Revisited}",
      journal = {\apj},
         year = 2021,
        month = may,
       volume = {912},
       number = {1},
          eid = {16},
        pages = {16},
          doi = {10.3847/1538-4357/abec44},
archivePrefix = {arXiv},
       eprint = {2103.02609},
 primaryClass = {astro-ph.SR},
       adsurl = {https://ui.adsabs.harvard.edu/abs/2021ApJ...912...16B}
}

@ARTICLE{Clark2014,
       author = {{Clark}, J.~S. and {Ritchie}, B.~W. and {Najarro}, F. and {Langer}, N. and {Negueruela}, I.},
        title = "{A VLT/FLAMES survey for massive binaries in Westerlund 1. IV. Wd1-5 - binary product and a pre-supernova companion for the magnetar CXOU J1647-45?}",
      journal = {\aap},
         year = 2014,
        month = may,
       volume = {565},
          eid = {A90},
        pages = {A90},
          doi = {10.1051/0004-6361/201321771},
archivePrefix = {arXiv},
       eprint = {1405.3109},
 primaryClass = {astro-ph.SR},
       adsurl = {https://ui.adsabs.harvard.edu/abs/2014A&A...565A..90C}
}

@ARTICLE{Jin2025,
       author = {{Jin}, Harim and {Langer}, Norbert and {Ercolino}, Andrea and {de Mink}, Selma E.},
        title = "{A comprehensive grid of massive binary evolution models for the Galaxy: Surface properties of post-mass-transfer stars}",
      journal = {\aap},
         year = 2026,
        month = feb,
       volume = {707},
          eid = {A56},
        pages = {A56},
          doi = {10.1051/0004-6361/202558177},
archivePrefix = {arXiv},
       eprint = {2510.19965},
 primaryClass = {astro-ph.SR},
       adsurl = {https://ui.adsabs.harvard.edu/abs/2026A&A...707A..56J}
}

@ARTICLE{deMink2014,
       author = {{de Mink}, S.~E. and {Sana}, H. and {Langer}, N. and {Izzard}, R.~G. and {Schneider}, F.~R.~N.},
        title = "{The Incidence of Stellar Mergers and Mass Gainers among Massive Stars}",
      journal = {\apj},
         year = 2014,
        month = feb,
       volume = {782},
       number = {1},
          eid = {7},
        pages = {7},
          doi = {10.1088/0004-637X/782/1/7},
archivePrefix = {arXiv},
       eprint = {1312.3650},
 primaryClass = {astro-ph.SR},
       adsurl = {https://ui.adsabs.harvard.edu/abs/2014ApJ...782....7D}
}

@ARTICLE{Eldridge2006,
       author = {{Eldridge}, J.~J. and {Genet}, F. and {Daigne}, F. and {Mochkovitch}, R.},
        title = "{The circumstellar environment of Wolf-Rayet stars and gamma-ray burst afterglows}",
      journal = {\mnras},
         year = 2006,
        month = mar,
       volume = {367},
       number = {1},
        pages = {186-200},
          doi = {10.1111/j.1365-2966.2005.09938.x},
archivePrefix = {arXiv},
       eprint = {astro-ph/0509749},
 primaryClass = {astro-ph},
       adsurl = {https://ui.adsabs.harvard.edu/abs/2006MNRAS.367..186E}
}

@ARTICLE{Scherbak2025,
       author = {{Scherbak}, Peter and {Lu}, Wenbin and {Fuller}, Jim},
        title = "{Rapid Binary Mass Transfer: Circumbinary Outflows and Angular Momentum Losses}",
      journal = {\apj},
         year = 2025,
        month = sep,
       volume = {990},
       number = {2},
          eid = {172},
        pages = {172},
          doi = {10.3847/1538-4357/adf067},
archivePrefix = {arXiv},
       eprint = {2505.21264},
 primaryClass = {astro-ph.SR},
       adsurl = {https://ui.adsabs.harvard.edu/abs/2025ApJ...990..172S}
}

@ARTICLE{Jin2024,
       author = {{Jin}, Harim and {Langer}, Norbert and {Lennon}, Daniel J. and {Proffitt}, Charles R.},
        title = "{Boron depletion in Galactic early B-type stars reveals two different main sequence star populations}",
      journal = {\aap},
         year = 2024,
        month = oct,
       volume = {690},
          eid = {A135},
        pages = {A135},
          doi = {10.1051/0004-6361/202450896},
archivePrefix = {arXiv},
       eprint = {2405.18266},
 primaryClass = {astro-ph.SR},
       adsurl = {https://ui.adsabs.harvard.edu/abs/2024A&A...690A.135J}
}

\appendix

\section{Hydrodynamic model}
\label{sec:App_hydro}

We performed our simulations using the (magneto-)hydrodynamic code \textsc{pion} \citep{2021PION}. Unless otherwise noted, the simulation setup and parameters used are as for the simulation presented in \citet{Larkin2025a}. In particular, we assume full photoionization due to the intense extreme-ultraviolet (EUV) radiation field ionising cool circumstellar material near the cluster core \citep{2018MNRAS.477L..55A}, include radiative heating and cooling with model 8 of \textsc{pion} as described in \citet{Green2019}, and do not include a magnetic field. We use six levels of static mesh refinement for the simulation, using a 2D cylindrical domain $D(r,z)$ of $r\in[0, 10.0\times 10^{18}]$ cm, $z\in[-10.0\times 10^{18}, 10.0\times 10^{18}]$ cm. $D$ is fully covered at level $n=0$ and each subsequent level of refinement covers a sub-domain of diameter $D/2^n$. We use $256\times512$ grid cells per level of refinement, such that the finest grid resolution is $\Delta x = 1.22\times10^{15}$ cm.

We assume a constant cluster wind of $\varv_z = -1500\kms$, density $ 1\times10^{-24}\mathrm{~g~cm}^{-3}$ and pressure $5\times 10^{-9} \mathrm{~dyn~cm}^{-2}$ \citep{Larkin2025b,Larkin2025a}. We use the MESA evolutionary track described above as input for the source term in our hydrodynamic model, with a wind injection radius of $5\times10^{16}$ cm ($\sim 20$ cells). We begin the simulation at 6.779 Myr to establish a pre-RLOF stellar bubble, and run it until 6.82 Myr, by which time the RLOF material has left the domain. We use the \mdot\ value of the primary star throughout the simulation, assuming fully non-conservative mass transfer. We calculate \vinf\ from the MESA track by scaling the primary star's escape velocity by the \teff-dependent prescription of \citet{Eldridge2006} except for during RLOF where we use a fixed value of 20\kms\ \citep{Scherbak2025}.

\section{MESA track}
\label{sec:App_evol}
Here we address the compatibility of our assumed MESA track with the known age, luminosity, period and surface H abundance of Wd1-72. Our track is 6.8~Myr post-ZAMS which is comparable to the cluster age of $\sim4.5-6.5$~Myr found by \citet{Castellanos2026}. \citet{Rosslowe2015} derive a luminosity of $\log L=5.45$ for the WR star. Our MESA track has a value of $\log L=5.0$ for $t=6.80$ Myr, but this difference can be accounted for by the choice of extinction law in \citet{Rosslowe2015} and the poorly constrained contribution of the secondary star. Our MESA track has $P=6.6$d at $t=6.80$ Myr, and reaches 6.9d at $t=6.81$ Myr. At $t=6.80$ Myr in the MESA track, surface H abundance is less than 10\%, with depletion to 1\% being reached at $t\sim6.805$ Myr.

There is a mild discrepancy between the time when the droplet morphology corresponds well with the observations ($t\sim6.80$ Myr) and when the MESA track orbital period and surface H abundance best match the measured properties of Wd1-72 ($t\sim6.81$ Myr). At this later time we observe the droplets are blown out of the simulation domain by the cluster wind. 

We also show two other simulations for comparison. In Fig. \ref{fig:app_def} we show the result of using the default MESA track without enhanced \mdot. The RLOF material fragmentation is slower, and we see the material is partially sheared by the cluster wind. To test the sensitivity to Mach number in Fig. \ref{fig:app_1000} we show the result of using the enhanced MESA track, but with a cluster wind velocity of 1000\kms. Qualitatively the morphology is similar, with the clumps dispersing slightly faster in this simulation. In both cases the key morphological properties are reproduced.

\begin{figure}
    \centering
    \includegraphics[width=1\linewidth]{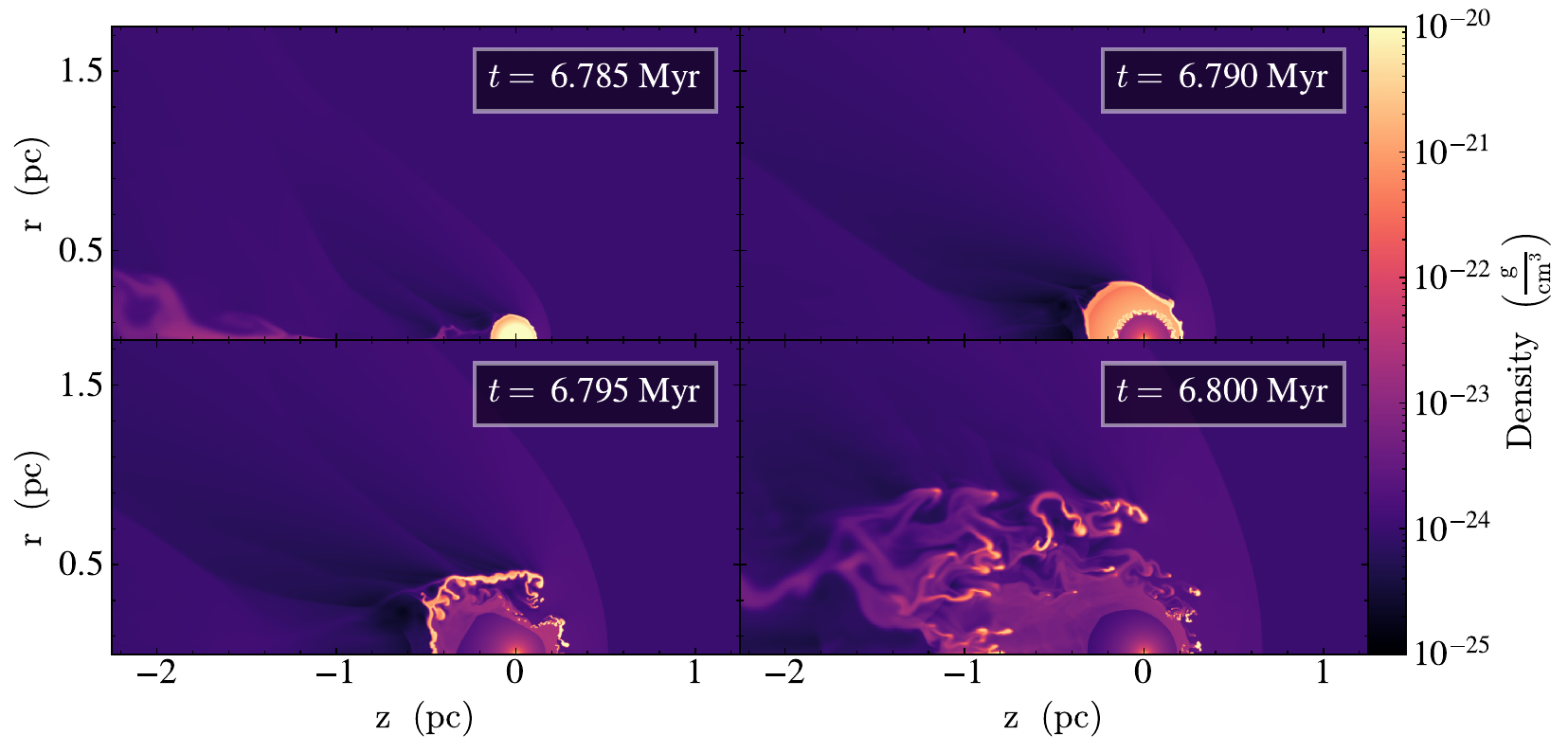}
    \caption{Density slice similar to the right panel of Fig.\,\ref{fig:mdot_hydro} but employing the default MESA track in the simulation.}
    \label{fig:app_def}
\end{figure}

\begin{figure}
    \centering
    \includegraphics[width=1\linewidth]{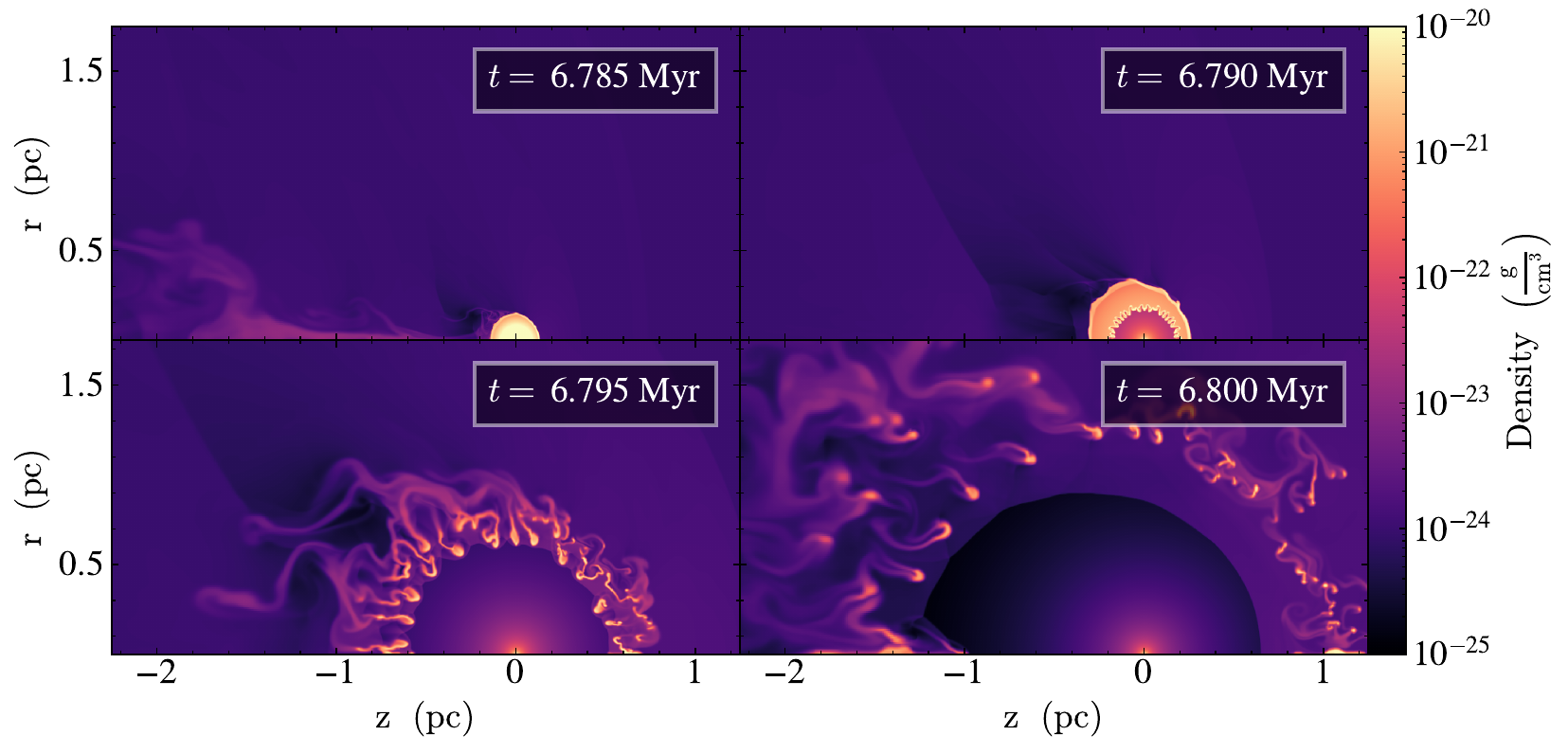}
    \caption{Density slice similar to the right panel of Fig.\,\ref{fig:mdot_hydro} but using a cluster wind velocity of 1000\kms\ in the simulation.}
    \label{fig:app_1000}
\end{figure}

\section{Physical constraints on the eruption and droplet properties}
\label{sec:App_analytic}
Here we demonstrate analytically that a giant LBV eruption cannot account for the observations around Wd1-72. Assuming an eruption instead of RLOF ML, the fragmentation can only occur when the ram-pressures of the eruption and cluster wind are approximately balanced, creating a standoff bow shock. RT fingers would then develop in the wake of this bow shock. From the data we see the standoff distance would be at $d\sim0.5$~pc\ from the star. We can balance the ram-pressures to obtain the ejection wind density $\rho_E$ at $d$:

\begin{multline*}
    \rho_E = 2.5\times10^{-23}\cdot\left(\frac{\rho_{CW}}{10^{-24}~\mathrm{g~cm}^{-3}}\right)\left(\frac{\varv_{CW}}{10^3 ~\mathrm{km~s}^{-1}}\right)^2 \\ \left(\frac{\varv_{E}}{200 ~\mathrm{km~s}^{-1}}\right)^{-2} ~\mathrm{g~cm}^{-3}
\end{multline*}

Where $\rho_{CW}, \varv_{CW}$ and $\varv_{E}$ are the cluster wind density, cluster wind velocity and eruption velocity respectively. Applying the continuity equation, the equivalent eruption \mdot\ for this density can be expressed as:

\begin{multline*}
    \dot{M}_E = 4\pi d \rho_E \varv_{E} = 2.37\times 10^{-4} \left(\frac{\rho_{CW}}{10^{-24}~\mathrm{g~cm}^{-3}}\right) \left(\frac{\varv_{CW}}{10^3 ~\mathrm{km~s}^{-1}}\right)^2 \\ \left(\frac{\varv_{E}}{200 ~\mathrm{km~s}^{-1}}\right)^{-1} \left(\frac{d}{0.5~\mathrm{pc}}\right) \Msun\pyr
\end{multline*}
    
For reasonable choices of these parameters, it is clear \mdot\ must be of order a few $10^{-4}$ \Msun\pyr\ at most for the ejecta to fragment in the cluster wind at this distance, assuming $\varv_E\sim$200\kms. Higher \mdot\ is only possible if $\varv_E$ is around an order of magnitude less, such as for RLOF. In this case, a dense shell of CSM develops during the RLOF phase, which is then fragmented from within via RT instabilities driven by the WR wind. We see this mechanism operating in Fig. \ref{fig:Vsl_app}, where we show a slice of $z-$axis velocity ($\varv_z$) from our simulation.

\begin{figure}
    \centering
    \includegraphics[width=1\linewidth]{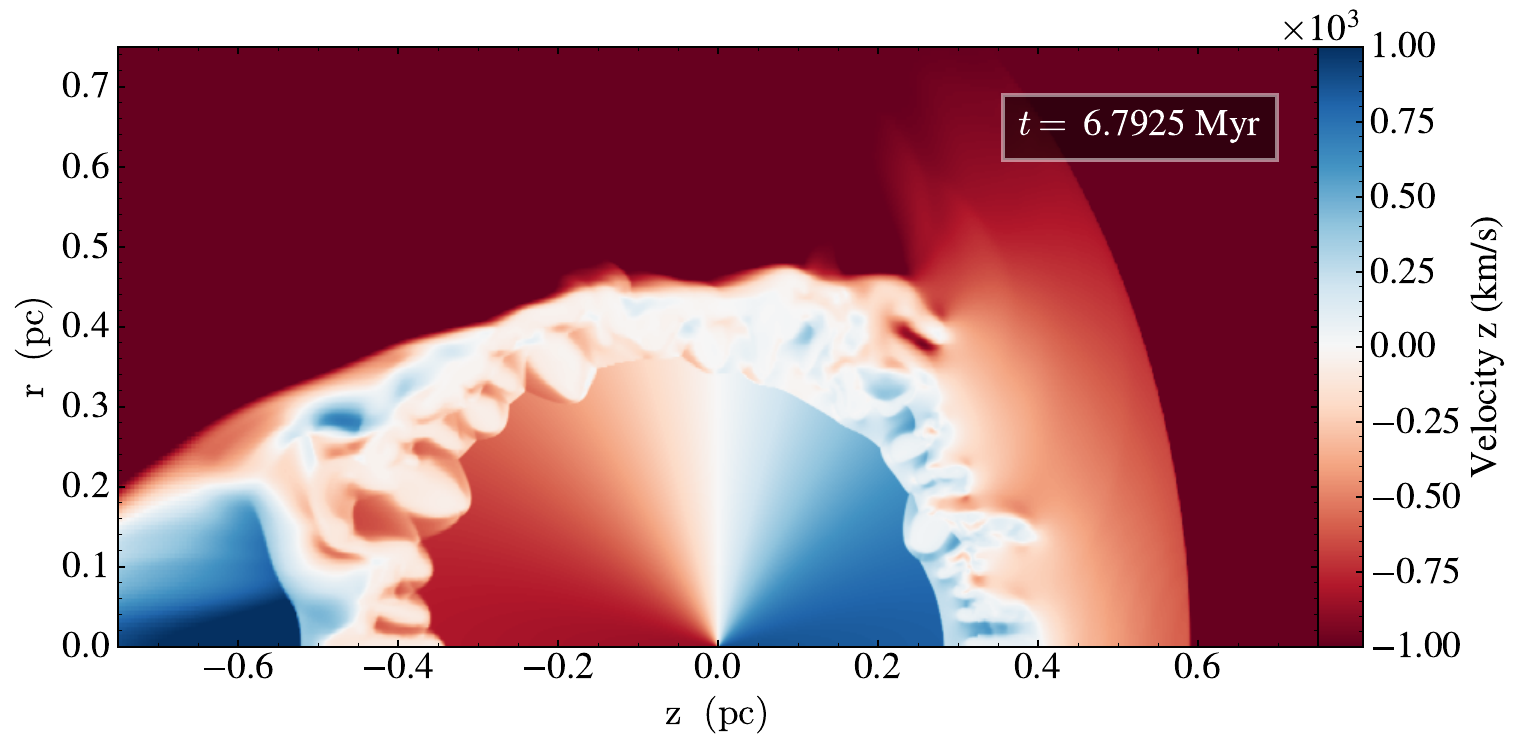}
    \caption{Simulation slice of $z-$axis velocity at 6.7925 Myr.}
    \label{fig:Vsl_app}
\end{figure} 

The droplets towards the cluster wind have expansion velocities $\varv_{\mathrm{exp}} = \sqrt{\varv_z^2 + \varv_r^2}$ of order $\lesssim 50\kms$ as they are at the stagnation point of the flow and so are decelerated. The downstream droplets are accelerated by the WR wind and/or cluster wind, whichever impacts them. We show a slice of $\varv_{\mathrm{exp}}$ in Fig. \ref{fig:vexp_app}.

\begin{figure}
    \centering
    \includegraphics[width=1\linewidth]{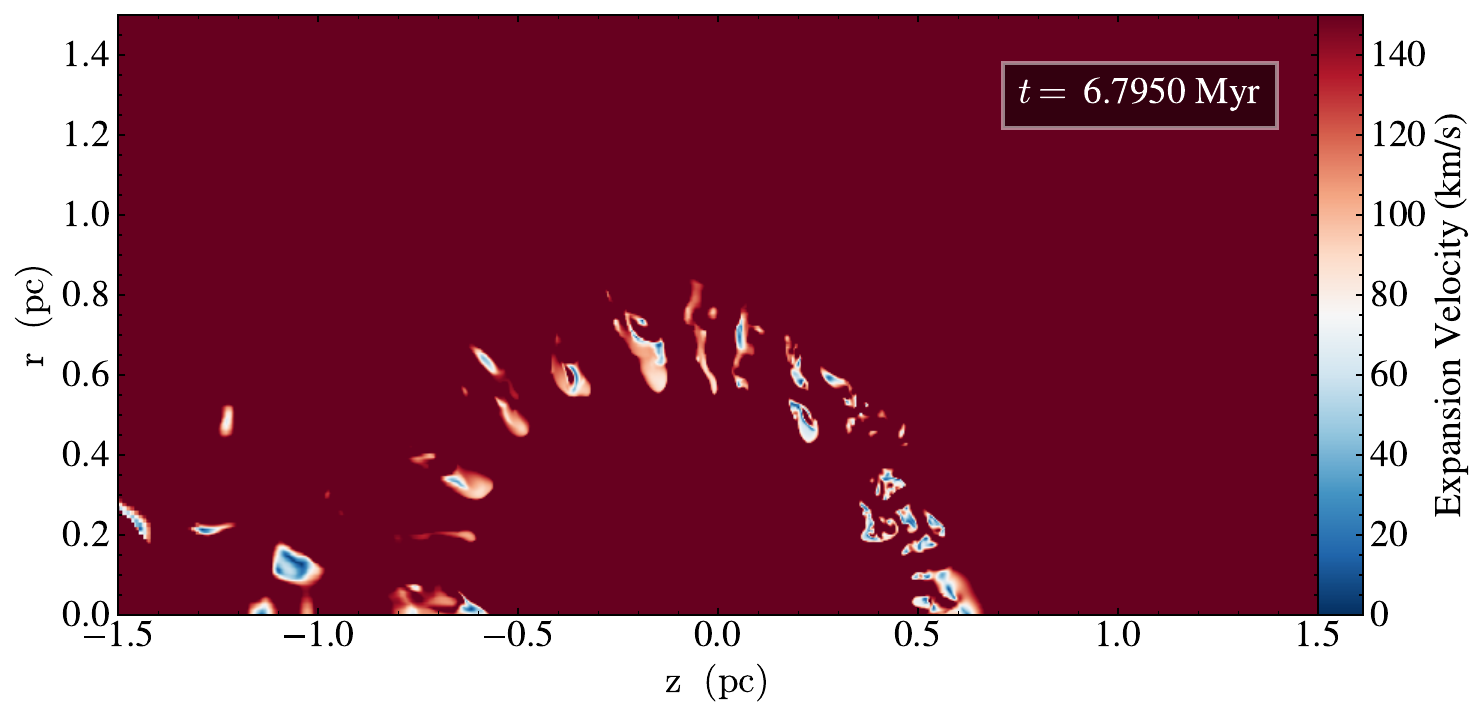}
    \caption{Simulation slice of expansion velocity at 6.795 Myr.}
    \label{fig:vexp_app}
\end{figure}

Fig.~\ref{fig:app:shear} shows again the $z$ component of velocity, $v_z$, but here at the later time of 6.800\,Myr.  Panels (b) and (c) show $v_z$ along almost-vertical lines passing through the head of the dense clumps in the downstream and upstream regions, respectively.
Apart from the velocity offset, we can see that the velocity shear in the downstream region is about $800\,\mathrm{km\,s}^{-1}$ between the heads of the dense clumps and the hot gas streams past them.
In contrast, the upstream region has much less uniform shear that varies from $150-500\,\mathrm{km\,s}^{-1}$.
The stagnation of the flow in the upstream region due to external ram pressure explains why the `droplets' do not have coherent cometary tails when compared with the downstream region.
In 2D simulations the flow is artificially constrained by the rotational symmetry, and it is likely that 3D simulations would have even more random flow patterns in the upstream direction.
The lack of external pressure in the downstream direction allows a more ordered flow to develop, producing cometary globules with coherent tails pointing away from Wd1-72.

\begin{figure}
    \centering
    \includegraphics[width=0.9\linewidth]{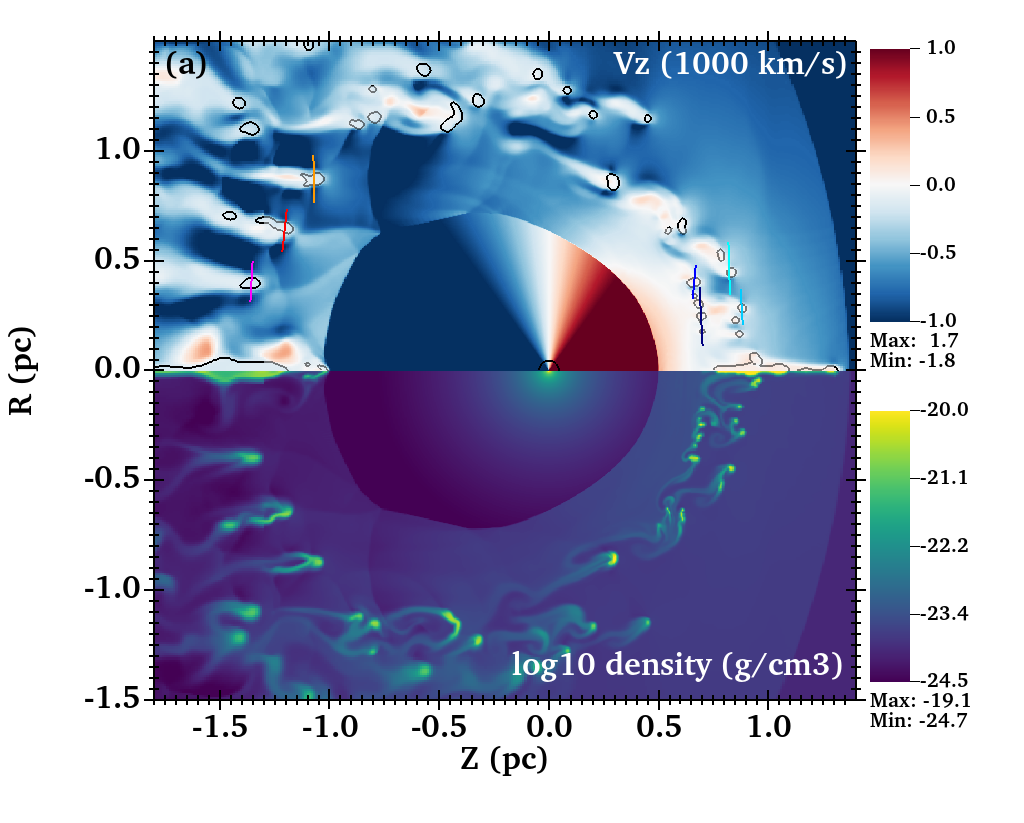}
    \includegraphics[width=0.9\linewidth, trim={0cm 1cm 0 1cm}, clip]{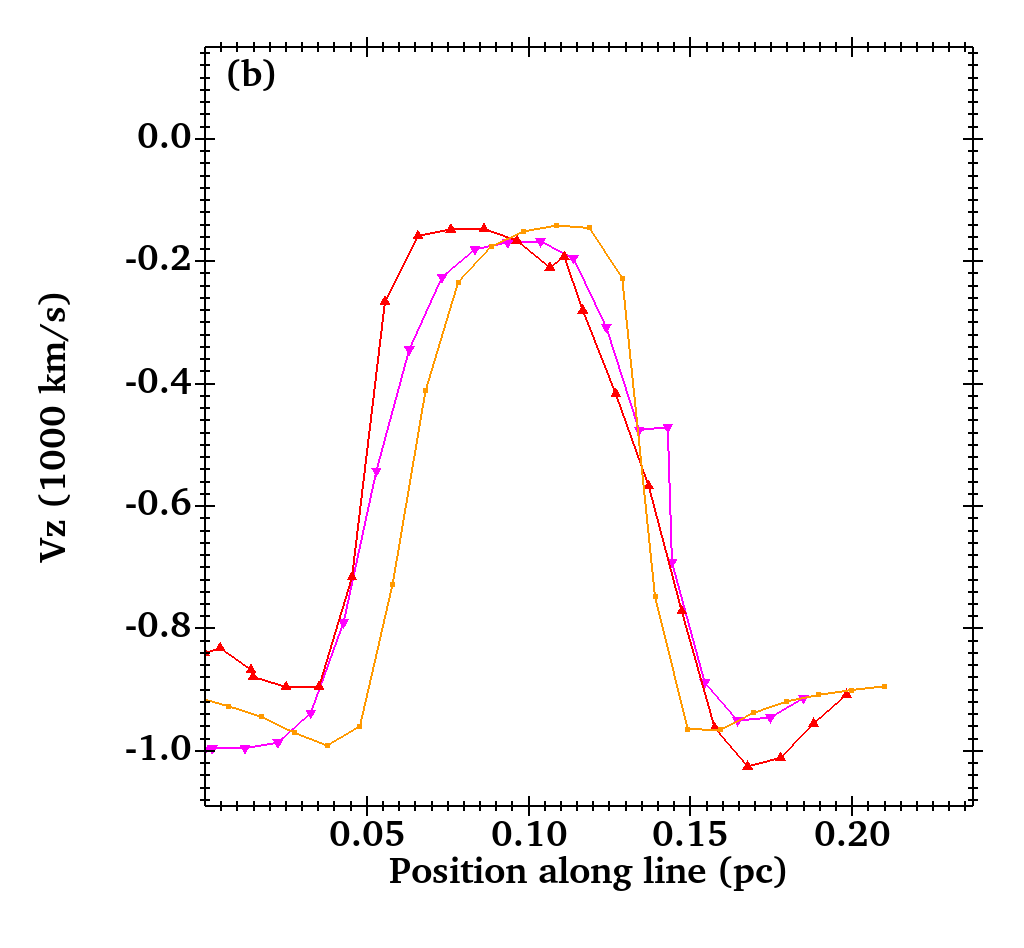}
    \includegraphics[width=0.9\linewidth, trim={0cm 1cm 0 1cm}, clip]{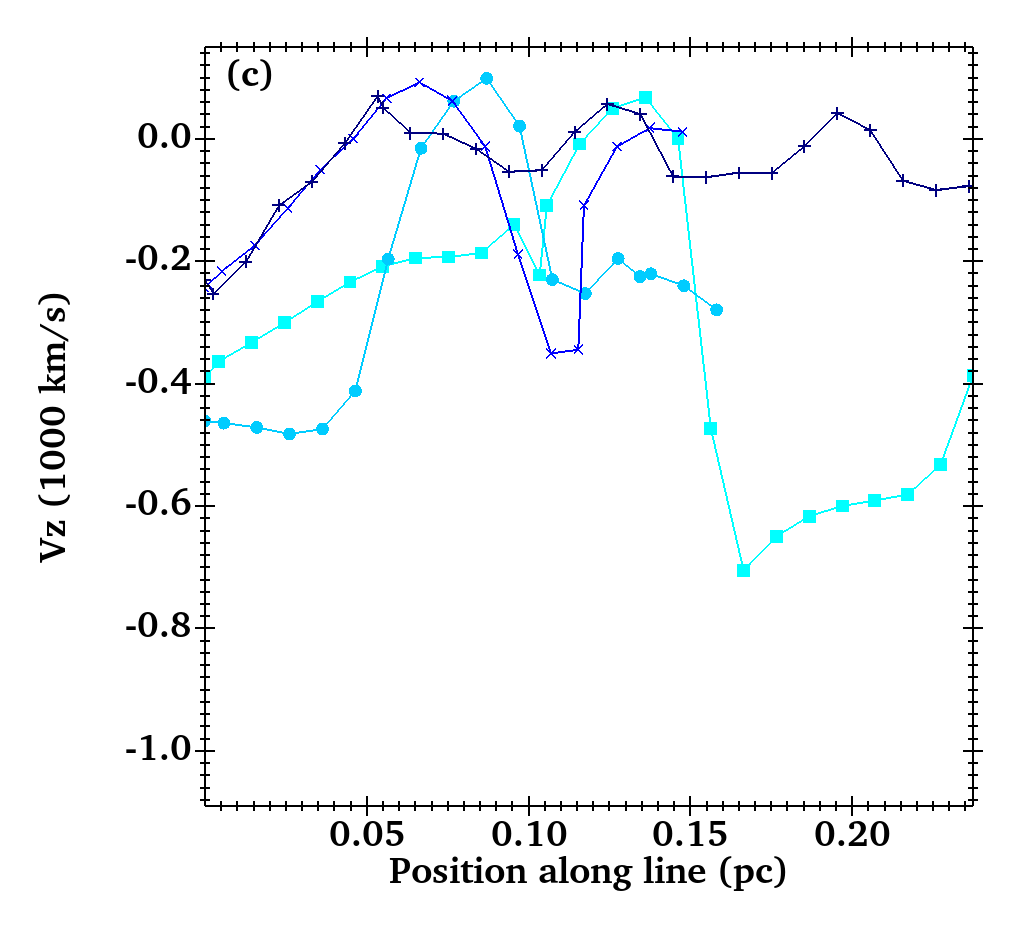}
    \caption{
        Velocity shear across dense clumps at $t=6.800\,Myr$: (a) horizontal component of the gas velocity, $v_z$, is shown in the upper half-plane and gas density ($\log_{10}\rho / \mathrm{g\,cm}^{-3}$) in the lower half-plane.
        Dense clumps with $\rho>10^{-22}$\,g\,cm$^{-3}$ are indicated by the black contours, and the seven almost-vertical lines show the slices where the $v_z$ is plotted in panels (b) and (c).
        Panel (b) shows $v_z$ in the colour-coded lines across the three downstream dense clumps, with distance measured from the large-$z$ end of the line, to show the velocity shear at the edge of the clumps.
        Panel (c) shows the same but for the upstream dense clumps.
        }
    \label{fig:app:shear}
\end{figure}

\section{Dust emission from dense clumps}
\label{sec:App_torus}

We used the \textsc{TORUS} Monte Carlo radiative transfer code \citep{Harries2019} to calculate thermal dust emissivity and dust temperature in the droplets produced by our simulation. 
We assume a constant 1\% dust mass-fraction for the RLOF material, and that the fast stellar winds and the cluster wind are dust free.
We use the same TORUS settings as \citet{Larkin2025a}, specifically assuming  silicate grains \citep{Dra03} and a grain-size distribution with power-law index of -3.3 from 0.005 to 0.2\,$\mu$m. 

We added six radiation sources, all of which are at $R=0$ along the $z$-axis, imposed because of symmetry constraints of the 2D simulations.
Wd1-72 is treated as a source with $\log_{10} L / \mathrm{L}_\odot = 5.45$ and 65\,kK at $z=0$\,pc,
the core of Wd~1 with   $\log_{10} L / \mathrm{L}_\odot = 5.9$, $T=40$\,kK at $z=1$\,pc,
WR\,U with $\log_{10} L / \mathrm{L}_\odot = 5.5$, $T=38$\,kK at $z=0.5$\,pc,
two sources representing nearby (in projection) WR stars with $\log_{10} L / \mathrm{L}_\odot = 5.2$, $T=50$\,kK at $z=0.48$ and 0.52\,pc,
and a source in the downstream region with $\log_{10} L / \mathrm{L}_\odot = 5.5$, $T=7.5$\,kK at $z=-0.25$\,pc.

\begin{figure}
    \centering
    \includegraphics[width=\columnwidth, trim={2.5cm 0cm 0 0cm}, clip]{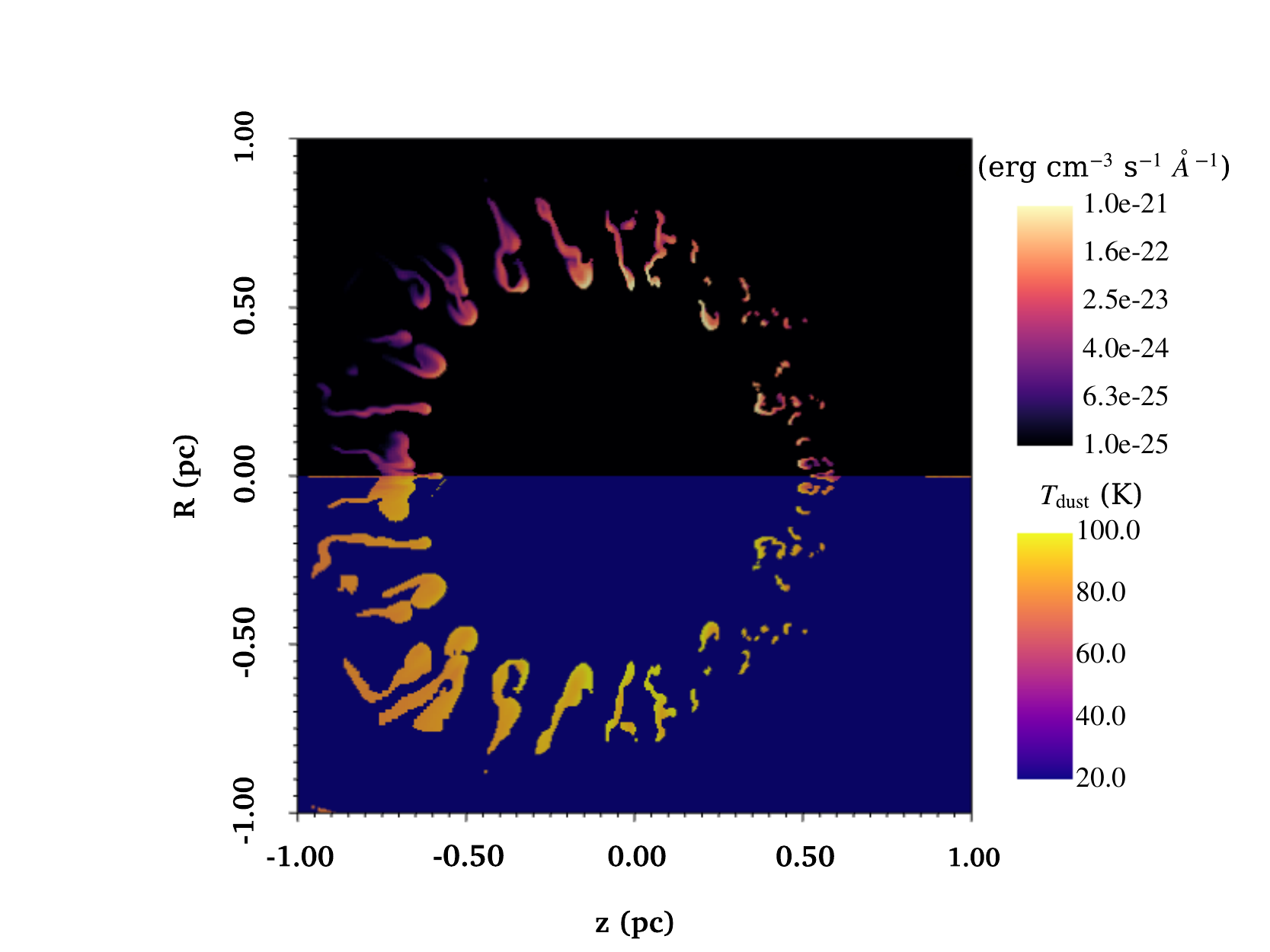}
    \caption{Upper half plane: thermal dust emissivity ($\epsilon$) at at $t = 6.795$ Myr calculated with the \textsc{torus} radiative transfer code, calculated at $\lambda=11\,\mu$m in units of erg\,cm$^{-3}$\,s$^{-1}$\,$\AA^{-1}$.
    Lower half-plane: dust temperature ($T_\mathrm{dust}$) assuming radiative equilibrium, where blue regions are dust-free and so the temperature is not defined.
    }
    \label{fig:dustemission}
\end{figure}  

Results of the \textsc{torus} modelling are shown in Fig. \ref{fig:dustemission}, where we plot dust emissivity $\epsilon$ (the \textsc{torus} variable \texttt{etacont} with units erg\,cm$^{-3}$\,s$^{-1}$\,$\AA^{-1}$) in the upper half-plane and dust temperature in the lower half-plane.
We did not calculate emission maps because the rotational symmetry imposed by the 2D simulation converts all features into rings.

Looking at $T_\mathrm{dust}$, the droplets towards the core of Wd~1 are warmer than those in the downstream direction, as expected due to the intense cluster radiation field.
The tails of the globules (at $z<-0.5$\,pc) are even cooler and their emission at $11\,\mu$m is faint in comparison to the rest of the dusty material.
Comparing with Fig.~\ref{fig:morphologycomparison} the same morphology is seen in the JWST image.
We cannot quantitatively compare the brightness of simulated dust emission with observations without a 3D simulation, which is currently too computationally expensive to justify given the large uncertainties in the mass and expansion velocity of the droplets and globules.
 
\end{document}